\documentclass[prb,twocolumn,showpacs,preprintnumbers,amsmath,amssymb]{revtex4} 
\usepackage{graphicx}% Include figure files%

\begin{document}
 
\title{DFT calculation of the intermolecular exchange interaction in 
the magnetic Mn$_4$ dimer}
\author{Kyungwha Park$^{1,2,}$}\email{park@dave.nrl.navy.mil}
\author{Mark R. Pederson$^{1}$}
\author{Steven L. Richardson$^{1,2}$}
\author{Nuria Aliaga-Alcalde$^{3}$}
\author{George Christou$^3$}
\affiliation{
$^1$Center for Computational Materials Science, Code 6390, 
Naval Research Laboratory, Washington DC 20375 \\
$^2$Department of Electrical Engineering and Materials Science Research Center, 
Howard University, Washington DC 20059 \\
$^3$Department of Chemistry, University of Florida, Gainesville, 
Florida 32611}
\date{\today} 
 
\begin{abstract}
The dimeric form of the single-molecule magnet [Mn$_4$O$_3$Cl$_4$(O$_2$CEt)$_3$(py)$_3$]$_2$
recently revealed interesting phenomena: no quantum tunneling at zero
field and tunneling before magnetic field reversal. This is attributed to 
substantial antiferromagnetic exchange interaction between different monomers. 
The intermolecular exchange interaction, electronic structure and 
magnetic properties of this molecular magnet are calculated using density-functional theory
within generalized-gradient approximation. Calculations are in good agreement
with experiment.

\end{abstract}

\maketitle

%\begin{multicols} 

%\section{Introduction}

Single-molecule magnets (SMMs), such as  
[Mn$_{12}$O$_{12}$(CH$_3$COO)$_{16}$(H$_2$O)$_4$]$\cdot$2(CH$_3$COOH)$\cdot$4(H$_2$O)
(hereafter Mn$_{12}$)\cite{LIS80} and  
[Fe$_8$O$_2$(OH)$_{12}$(tacn)$_6$]Br$_8$$\cdot$9(H$_2$O) (hereafter Fe$_8$)\cite{WIEG84}
have received tremendous attention due to macroscopic quantum tunneling\cite{VILL94}
and possible use as nanomagnetic storage devices. 
Hysteresis loop measurements on the SMMs Mn$_{12}$ and Fe$_8$ showed 
magnetization steps at low temperatures upon magnetic field reversal.\cite{SESS93}
This is due to quantum tunneling between spin-up states and spin-down states 
despite a large effective spin $S$$=$10 for each molecule. The
resonant tunneling fields in these systems are primarily determined
by the magnetomolecular anisotropy. Recently a dimerized 
single-molecule magnet [Mn$_4$O$_3$Cl$_4$(O$_2$CEt)$_3$(py)$_3$]$_2$
(hereafter Mn$_4$ dimer) where Et=CH$_2$CH$_3$ and py=NC$_5$H$_5$,
has been formed\cite{HEND92,WERN02-NAT} which exhibited qualitatively different tunneling 
behavior: quantum tunneling prior to magnetic field reversal 
and an absence of quantum tunneling at zero field 
in contrast to other SMMs such as Mn$_{12}$ and Fe$_8$.\cite{WERN02-NAT}
To understand the basis for the qualitative deviation we
have calculated both the magnetomolecular anisotropy and the intermolecular
exchange interaction in the Mn$_4$ dimer using density-functional theory. 
Our results confirm that there exists an appreciable {\it antiferromagnetic} 
exchange interaction between monomers and that tunneling
fields in this dimer are strongly influenced by the presence
of the monomer-monomer exchange interaction. This interaction
produces a bias field that encourages
monomeric magnetic-moment reversal below zero field and 
prevents two monomers from simultaneously flipping 
their magnetic moments at zero field.  We determine that the origin of the
exchange interaction is not dominated by either kinetic or exchange-correlation
terms and that the total "exchange" interaction is in fact an order of magnitude
smaller than the kinetic contribution.
For Mn$_{12}$ and Fe$_8$, the intermolecular exchange interaction has not been 
observed experimentally and it is generally accepted that the overlap between
neighboring molecules is negligible.
 
\begin{figure}[tb]
\includegraphics[angle=0,width=0.27\textwidth]{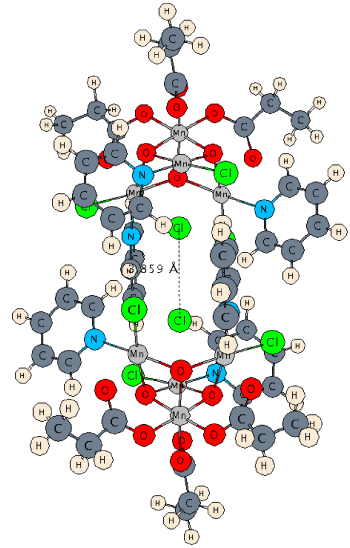}
%\begin{center}
%\epsfxsize=15.cm
%\epsfysize=20.cm
%\epsfbox{kyungwha_exp.eps}
\caption[]{Mn$_4$ dimer geometry. The dimer is formed by inversion
of the threefold symmetric monomer. Each monomer
has a magnetic core consisting of three ferromagnetically
coupled Mn$^{3+}$ spins ($S$$=$2) coupled antiferromagnetically to
one Mn$^{4+}$ spin ($S$$=$3/2) ion leading to a                                    
total spin of $S=9/2$.  The distance between the two central Cl atoms marked as the
dotted line was measured to be 3.86~\AA.~}
\label{fig:Mn4geo}
%\end{center}
\end{figure}

In this work, we discuss calculations on the Mn$_4$ dimer which is formed
by inversion of the three-fold symmetric monomer shown in Fig.\ref{fig:Mn4geo}. 
The magnetic core of the Mn$_4$ monomer consists of three 
ferromagnetically coupled Mn$^{3+}$ ($S$$=$2) ions coupled antiferromagnetically to
the remaining Mn$^{4+}$ ($S$$=$3/2) ion                               
leading to a total ground-state spin of 
$S$$=$2$\times$3$-$3/2$\times$1$=$9/2 (refer to Fig.\ref{fig:Mn4geo}).
The core has a similar cubane structure
as the inner core of the SMM Mn$_{12}$, although there are four 
Mn$^{4+}$ for Mn$_{12}$. 
We investigate the electronic structure and magnetic properties of 
this SMM Mn$_4$ using density-functional theory (DFT). We calculate optimized geometries 
for the Mn$_4$ monomer and dimer, their binding energy, the monomeric
magnetic anisotropy barrier (MAE), and the exchange coupling constant 
between monomers. Results are compared with experiment.

Our DFT calculations\cite{KOHN65} are performed with the all-electron 
Gaussian-orbital-based Naval Research Laboratory Molecular Orbital 
Library (NRLMOL).\cite{PEDE90} Here we use the Perdew-Burke-Ernzerhof (PBE) 
generalized-gradient approximation (GGA).\cite{PERD96} 
Before discussing energetics and magnetic phenomena we discuss two
structural issues. First, we have considered monomers and dimers that
are terminated by both H and by the CH$_2$CH$_3$ radicals found in the
experimental structure.
Second we have considered structures based on two conformers of the
monomeric unit. While a complete vibrational analysis will be discussed
in a later publication, all indications are that both conformers are
stable. The conformers have slightly different arrangements of the
pyridine ligands. The first conformer was identified by a
density-functional-based geometry optimization of the hydrogenated 
monomer.
The second conformer was identified by improvements on the monomer
deduced from the experimental x-ray data. In the remainder of the
paper we refer to these monomers as the computationally determined
conformer (CDC) and the experimentally determined conformer (EDC).
%Our DFT calculations find that the CDC monomer is higher in
%energy than the EDC monomer by 0.62 eV and that the dimer bonding energies
%of the EDC hydrogenated monomer is 0.45 eV compared to 0.1 eV for the
%CDC hydrogenated monomer. A second structural issue is related to the role
%of radical termination of the tails of the monmers. The dimer binding
%energies show a sensitivity to the H vs. CH$_2$CH$_3$ tail terminators 
%(0.45 vs 0.78 eV). As discussed below, these structural issues have 
%neglibible effect on the magnetic properties.

Each Mn$_4$ monomer has threefold symmetry so there are 
26 inequivalent atoms to consider. The number of inequivalent atoms 
is reduced to 20 when the CH$_2$CH$_3$ radical is replaced by H.
A pyridine ring is initially constructed to lie in the plane defined by the 
vector connecting Mn$^{3+}$ and neighboring N and the sum of the two vectors 
connecting Mn$^{3+}$ with two closest Cl's (refer to Fig.~\ref{fig:Mn4geo}). 
The geometries for the pyridine ring and the cubane were first optimized separately 
to generate an initial geometry for the DFT calculations on the full monomer.
The initial geometry for the monomer was relaxed using NRLMOL with the Cl atom 
fixed to reproduce the experimental Cl-Cl distance  (3.86~\AA)~upon 
dimerization (i.e. adding inversion symmetry).  Relaxation continues until forces 
exerted on all atoms become $\sim$0.001~hartree/bohr. 
The CDC dimer is then obtained by inversion of the CDC monomer
with the fixed value of $d=3.86$~\AA~(marked as dotted in Fig.~\ref{fig:Mn4geo}).
For the case of the x-ray deduced experimental geometry, the      
C-H bond lengths are underestimated (0.71~\AA~ to 0.96~\AA)~ in comparison to
standard hydrogen bond lengths, which yields self-consistent forces on hydrogen 
atoms  as large as 0.8~hartree/bohr.  To improve the experimental geometry, all 
hydrogen positions were 
first moved to create C-H bond lengths as 1.1~\AA, and then additional optimization
of the experimental geometry was performed with the fixed Cl-Cl distance.          
The experimental geometry without corrected hydrogen positions was             
53~eV higher in energy than that of the structure with corrected hydrogen positions.             
Hereafter unless we specify, the EDC monomer refers to the
optimized experimental geometry with corrected hydrogen positions.

\begin{figure}[tb]
\includegraphics[angle=0,width=0.32\textwidth]{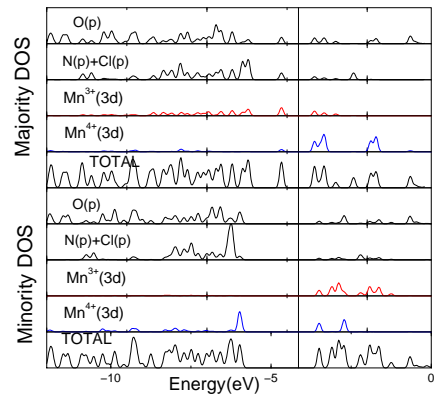}
%\epsfxsize=15.cm
%\epsfysize=20.cm
%\epsfbox{DOS_color_port.eps}
\caption{Electronic density of states (DOS) for majority and minority spins
for the Mn$_4$ monomer. Shown are projected Mn(3d) DOS of the two types of Mn ions,
projected p DOS of the three N atoms and the four Cl atoms, projected p DOS of the nine
O atoms, and the total DOS defined by the sum of projected DOS of all atoms in
the monomer. All projected DOS have the same scale which is different from that
for the total DOS. The vertical line denotes the Fermi level.
Directly below the Fermi level, for majority spins the projected Mn$^{3+}$(3d)
DOS has more weight than the Mn$^{4+}$(3d) DOS. For minority spins,
the tendency is the opposite.}
\label{fig:DOS}
\end{figure}

We have used full basis sets for all six different atoms and fine mesh.\cite{PORE99} 
Charges and magnetic moments for Mn's from the CDC monomer agree well with 
those from the EDC monomer.  For example, a sphere with a radius of 2.23 Bohr 
captures charges of 23.4 and 23.7, and magnetic moments of $3.6\mu_B$ and $-2.5\mu_B$ for Mn$^{3+}$ 
and Mn$^{4+}$ respectively. The total magnetic moment for the monomer 
is $9\mu_B$ in good agreement with experiment. The HOMO-LUMO gap 
for majority (minority) spin is 1.02~eV (2.42~eV). 
The energy difference between the minority (majority) 
LUMO and the majority (minority) HOMO is 1.17~eV (2.28~eV), which ensures
that the system is stable with respect to the total magnetic moment. 
%Figure~\ref{fig:DOS} shows the total electronic density of states (DOS), 
%the projected Mn(3$d$) DOS for the two types of Mn ions, the projected $p$ DOS of
%three N and four Cl atoms, and projected O($p$) DOS of nine O atoms
%for the majority and minority spin in the monomer. 
As clearly seen in Fig.~\ref{fig:DOS}, right below the 
Fermi level for majority spins the projected Mn$^{3+}$(3$d$) DOS 
is dominant over the projected Mn$^{4+}$(3$d$) DOS, while 
for minority spin the opposite trend is observed. This confirms
the experimental picture of three Mn$^{3+}$ spins 
antiferromagnetically coupled to a Mn$^{4+}$ spin.

We calculate the binding energy by subtracting the dimer energy 
from twice the monomer energy. We find that the dimer is stable
for both the CDC and EDC.  For the CDC (EDC), the binding energy is about 
0.16~eV (0.78~eV).  The magnitude of the binding energy suggests attractive
electrostatic interactions between different monomers. 
The discrepancy between the binding energy for the CDC and that for 
the EDC may be attributed to our substitution of ethyl for hydrogen in 
the CDC and/or the fact that the plane where a pyridine ring sits 
is different for both geometries.  
To check the former possibility, we calculate
the binding energy of the EDC terminated by hydrogen, and obtain 0.45~eV. 
We have also verified that the conformation of a pyridine ring for 
the EDC is slightly different from that for the CDC. Thus, the discrepancy 
arises from both reasons. 

We have calculated the monomeric MAE in zero magnetic 
field for both the CDC and EDC with the assumption
that spin-orbit coupling is a major contribution to the MAE. 
For this calculation, we follow the procedure developed in Ref.\onlinecite{PEDE99}. 
Our calculations show that the Mn$_4$ monomer has uniaxial anisotropy 
along the threefold axis (the bond between Mn$^{4+}$ and Cl in the 
cubane), in agreement with experiment.\cite{HEND92,WERN02-NAT} 
For uniaxial systems, the energy shift $\Delta$ due to
the spin-orbit interaction can be simplified to 
$-\gamma_{zz} \langle S_z \rangle^2$ up to constant terms
independent of $\langle S_z \rangle$ if the $z$ axis is assigned
as the easy axis.\cite{PEDE99} Then the classical barrier (MAE) 
to be overcome to monomer magnetization reversal 
$M_z$$=$+9/2 to $M_z$$=$$-$9/2 is $\gamma_{zz} ((9/2)^2-(1/2)^2)$.
For the CDC (EDC) monomer, the MAE is 11.3~K (11.6~K), which
is close to that for the hydrogenated EDC monomer. As shown in
Table~\ref{table:1}, all these numbers are close to the experimental value 
of 14.4~K. The difference between our estimated
MAE and the experimental value might be ascribed to other effects
on the barrier such as spin-vibron coupling.\cite{PEDE02} 

\begin{table}
\begin{center}
\caption{Binding energy, monomeric magnetic anisotropy barrier (MAE),
and antiferromagnetic exchange constant $J$ for the CDC
with the distance between the two central Cl's held as
the experimental value, $d=3.86$~\AA~[DFT(1)], the EDC
with $d=3.86$~\AA~[DFT(2)], and the same as DFT(2)
except that ethyl is replaced by hydrogen [DFT(3)].
DFT(4), DFT(5), and DFT(6) denote the same
as DFT(2) except that $d$$=$3.86~\AA$+$1~Bohr,
$d$$=$3.86~\AA$-$0.5~Bohr, and
$d$$=$3.86~\AA$-$1~Bohr, respectively.
The experimental values are from Ref.\onlinecite{WERN02-NAT}.
The numerical uncertainty in the estimated values of $J$ is $\sim 0.04$~K.}
\label{table:1}
\begin{ruledtabular}
\begin{tabular}{|c|c|c|c|}
 & Binding energy & MAE/monomer & exchange $J$ \\ \hline
DFT(1) & 0.16~eV & 11.3~K & 0.24~K \\ \hline
DFT(2) & 0.78~eV & 11.6~K & 0.27~K \\ \hline
DFT(3) & 0.45~eV & 10.9~K & \\ \hline
DFT(4) &  & 11.7~K & 0.10~K \\ \hline
DFT(5) &  & 11.6~K & 0.47~K \\ \hline
DFT(6) &  & 11.7~K & 0.81~K \\ \hline
Exp\cite{WERN02-NAT} &  & 14.4~K & 0.1~K \\
\end{tabular}
\end{ruledtabular}
\end{center}
\end{table}

To calculate the exchange coupling constant $J$ between monomers, we 
assume that a monomer is an ideal $S=9/2$ object and that its 
effective spin is aligned along the easy axis and of Ising type (either 
$M_z$$=$+9/2 or $-9/2$). Then we calculate self-consistently energies of 
ferromagnetic (parallel monomeric spins) and 
antiferromagnetic configuration (antiparallel monomeric spins) of the 
dimer, and take a difference $\delta$ 
between the two energies. We find that the antiferromagnetic 
configuration is favored.
The antiferromagnetic exchange constant $J$ is determined from 
$\delta=2J(9/2)^2$. For the CDC, the energy
difference is 31 microhartree so that $J$$=$0.24~K, 
while for the EDC, $J$$=$0.27~K. These can be compared to the experimentally
measured value of $J$$=$0.1~K.\cite{WERN02-NAT} 
The numerical uncertainty in the total-energy difference for our DFT calculations
is at most 5 microhartree, which can be translated 
to the uncertainty in the exchange $J$ as 0.04~K.
We achieve high-accuracy in the total-energy difference, because
we use exactly the same optimized dimer geometry with
the same parameter values for a self-consistent approximation
except for the effective spin configurations of monomers. Although
our DFT estimated value of $J$ is somewhat higher than
the experimental value, this may be acceptable considering
the assumptions we made and the fact that DFT calculations 
often overestimate exchange interactions.
In some cases, the PBE generalized-gradient approximation
may not fully cancel the self-interaction in the Coulomb potential.
Therefore, the electrons in our calculations are slightly
more diffuse, which should lead to overestimated exchange interaction.

\begin{figure}
\includegraphics[angle=0,width=.3\textwidth,height=.2\textwidth]{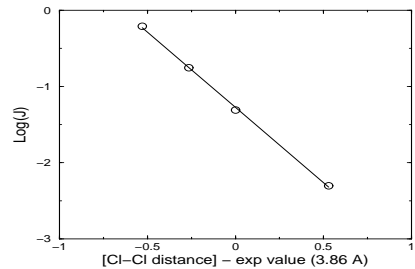}
%\epsfxsize=10.cm
%\epsfysize=8.cm
%\epsfbox{logJvsdist.eps}
\caption{Logarithm of exchange constant $J$ as a function of the
monomer-monomer distance relative to the experimental value.
The numerical uncertainty in $J$ is $\sim 0.04$~K.
The slope of the curve is about -2.}
\label{fig:logJvsdist}
\end{figure}

It is interesting to examine whether the exchange interaction
varies significantly with the monomer-monomer separation. 
We consider the case that each 
monomer is displaced toward or away from the center of mass 
of the dimer along the easy axis. Then we calculate 
the exchange constant $J$ for the EDC dimer 
with three different monomer-monomer distances
from the experimentally measured value. The monomer-monomer
distance is varied by changing the two central Cl-Cl distance
with a monomer geometry fixed. If the central Cl-Cl 
bond length increases by 1 Bohr, then $J$ decreases 
down to 0.10~K. If the bond length decreases by 0.5 Bohr (1 Bohr), 
$J$ increases to 0.47~K (0.81~K). Table~\ref{table:1}
summarizes the separation dependence of $J$ and of the monomeric MAE. 
As shown in Table~\ref{table:1}, the monomeric MAE does not depend on
the exchange interaction between monomers, because the 
monomer geometry has not changed during this process.
Figure~\ref{fig:logJvsdist} shows that
$J$ increases exponentially with decreasing
the separation distance. This tells us how quickly the 
overlaps of neighboring wavefunctions decrease with increasing 
the distance. We have decomposed the J values into kinetic, coulombic
and exchange-correlation contributions. The kinetic contribution is 
an order of magnitude larger than the total value of J and it is significantly
cancelled by the exchange-correlation contributions to the J value.

\begin{table}
\begin{center}
\caption{Initial $|M_1,M_2 \rangle$ and final states
$|M_1^{\prime},M_2^{\prime} \rangle$ participating in
quantum tunneling at resonant fields, $B_{\mathrm {res}}$. This
was calculated by exact diagonalization of Hamiltonian (\ref{eq:ham}).
$M_1$ and $M_2$ are the projected magnetic moment
along the easy axis for each monomer. For clarity, degeneracy in initial/final
states is not listed. Only for the case with $\gamma_{zz}\gg J$
the initial/final states are eigenstates of
Hamiltonian~(\ref{eq:ham}). The third and fourth
resonances are from one degenerate state to another degenerate
state, and they are split due to transverse
terms in the exchange interaction. The same logic is applied to
the last two resonances. B$_{\mathrm{res}}^{\mathrm{Exp}}$
is the resonant field for $\gamma_{zz}=0.72$~K and $J=0.1$~K.
B$_{\mathrm{res}}^{\mathrm{DFT}}$ is for
$\gamma_{zz}=0.58$~K and $J=0.27$~K
B$_{\mathrm{res}}^{\mathrm{Mod}}$ is for
$\gamma_{zz}=0.58$~K and $J=0.1$~K.}
\label{table:2}
\begin{ruledtabular}
\begin{tabular}{|c|c|c|c|c|}
Initial & Final & B$_{\mathrm{res}}^{\mathrm{Exp}}$(T) &
 B$_{\mathrm{res}}^{\mathrm{DFT}}$ &
B$_{\mathrm{res}}^{\mathrm{Mod}}$   \\ \hline
$|\frac{9}{2},\frac{9}{2} \rangle$ & $|\frac{9}{2},-\frac{9}{2} \rangle$
& $-0.335$ & $-0.915$ &  $-0.335$  \\ \hline
$|\frac{9}{2},\frac{9}{2} \rangle$ & $|\frac{9}{2},-\frac{7}{2} \rangle$
& 0.20 & $-0.495$ & 0.095  \\ \hline
$|\frac{9}{2},-\frac{7}{2} \rangle$ &
$|-\frac{9}{2},-\frac{7}{2} \rangle$ &
0.23 & 0.625 & 0.23  \\ \hline
$|\frac{9}{2},-\frac{7}{2} \rangle$  &
$|-\frac{9}{2},-\frac{7}{2} \rangle$ &
0.305 & 0.83 & 0.305  \\ \hline
$|\frac{9}{2},-\frac{9}{2} \rangle$  &
$|-\frac{9}{2},-\frac{9}{2} \rangle$ & 0.34 & 0.92 & 0.34  \\ \hline
$|\frac{9}{2},\frac{9}{2} \rangle$ & $|\frac{9}{2},-\frac{5}{2} \rangle$
& 0.735 & $-0.08$ & 0.525   \\ \hline
$|\frac{9}{2},-\frac{9}{2} \rangle$ &
$|-\frac{9}{2},-\frac{7}{2} \rangle$ &
0.835  & 1.24 & 0.73  \\ \hline
$|\frac{9}{2},-\frac{9}{2} \rangle$ &
$|-\frac{9}{2},-\frac{7}{2} \rangle$ &
0.915 & 1.465 & 0.815  \\
\end{tabular}
\end{ruledtabular}
\end{center}
\end{table}

Since we estimated the anisotropy barrier and exchange constant,
we can construct a model Hamiltonian for the dimer according to
\begin{eqnarray}
{\cal H}&=& -\gamma_{zz} (S_{1z}^2 + S_{2z}^2) + 
%{\cal H}_1^{\mathrm{tr}} + {\cal H}_2^{\mathrm{tr}} +
J \vec{S}_1 \cdot \vec{S}_2
\label{eq:ham}
\end{eqnarray}
where the uniaxial anisotropy parameter $\gamma_{zz}=0.58$~K 
and $J=0.27$~K. To determine whether our values of $\gamma_{zz}$
and $J$ can reproduce the experimental values of the resonant 
tunneling fields (Fig.~4 in Ref.\onlinecite{WERN02-NAT}),
we calculate these fields using exact diagonalization
of Hamiltonian~(\ref{eq:ham}). Although it is crucial 
to include some small transverse terms in the Hamiltonian
(such as transverse anisotropy and transverse fields)
for calculations of tunnel splittings, the transverse terms
 do not affect the resonant fields much.
Table~\ref{table:2} summarizes the resonant fields
for some low-energy states for three different values of
$\gamma_{zz}$ and $J$: (1) the experimental values 
$\gamma_{zz}=0.72$~K, $J=0.1$~K; (2) $\gamma_{zz}=0.58$~K,
$J=0.27$~K; (3) slightly modified version of our DFT results 
$\gamma_{zz}=0.58$~K, $J=0.1$~K. Let us focus on two tunnelings which
were prominent in the experimental measurements:
$|M_1=9/2,M_2=9/2 \rangle \rightarrow |M_1=9/2,M_2=-9/2 \rangle$, 
and $|9/2,-9/2 \rangle \rightarrow |-9/2,-9/2 \rangle$, where
$M_1$ and $M_2$ are the eigenvalues of the spin operator projected
along the easy axis for each monomer.
For these two tunnelings, the resonant fields are solely determined by $J$ 
and are independent of $\gamma_{zz}$: $B_{\mathrm{res}}\approx\mp9J/(2g\mu_B)$.
Therefore, model Hamiltonian~(\ref{eq:ham}) with our estimated values 
will not quantitatively reproduce the experimental resonant fields. 
However, in this case (when $\gamma_{zz}$ becomes comparable to $J$),
we notice that the hysteresis loop exhibits richer features such as more
magnetization steps before magnetic field reversal. 
Since DFT often overestimates exchange
interactions, we also calculate the resonant fields with
$J$ decreased to 0.1~K and $\gamma_{zz}$ fixed to examine 
if agreement with experiment improves.  We find that some resonances
agree with experiment and some do not agree.

In summary, we have calculated optimized geometries for a monomer and dimer of
the SMM Mn$_4$ using DFT. For both the CDC and EDC, 
we calculated binding energy, monomeric MAE, and the exchange 
interaction between monomers. The binding interaction between monomers 
is electrostatic. Our calculated anisotropy
barrier is close to the experimental value. The exchange interaction 
between monomers is twice or three times larger than the experimental value. 
Overall, our DFT calculations are in qualitative accord
with experiment.

%\section{Conclusion}

%\begin{center}
%{\textbf{Acknowledgments}}
%\end{center}
%\section*{Acknowledgments}
KP and SLR were funded by W. M. Keck Foundation, MRP was supported 
in part by ONR and the DoD HPC CHSSI program, and NA and GC were supported by 
NSF Grants Nos. CHE-0123603 and DMR-0103290.

%\end{multicols} 

\end{document}